\documentclass[%
nofootinbib,
 amsmath,amssymb,
 aps,
 prd,
  11pt,
tightenlines
]{revtex4-2}

\usepackage{graphicx}% Include figure files
\usepackage{dcolumn}% Align table columns on decimal point
\usepackage{bm}% bold math
\usepackage[dvipsnames]{xcolor}
\usepackage{amsmath}
\usepackage[english]{babel}
\usepackage[autostyle, english = american]{csquotes}
\MakeOuterQuote{"}
\usepackage{hyperref}

\usepackage[normalem]{ulem}
\begin{document}

\preprint{APS/123-QED}

%\title{Is there a connection between superfluid and Bose--Einstein condensate dark matter?}

\title{Bridging Superfluid and Nonminimally Coupled BEC Dark Matter\\ through RAQUAL}
% Force line breaks with \\
\author{Stefano Liberati}%
\email{liberati@sissa.it}

\author{Valentin Pomakov}
 \email{vpomakov@sissa.it}

\author{Samuele Silveravalle}
\email{ssilvera@sissa.it}

\affiliation{%
 SISSA, Via Bonomea 265, 34136 Trieste, Italy \\ INFN, Sezione di Trieste, Via Valerio 2, 34127 Trieste, Italy \\ IFPU--Institute for Fundamental Physics of the Universe, Via Beirut 2, 34014 Trieste, Italy
}%

\date{\today}% It is always \today, today,

\begin{abstract}
% Motivated by conceptual similarities based on analogies with condensed-matter systems, we look for a mapping between two modified-gravity dark matter models: Superfluid Dark Matter (SFDM) by Khoury and collaborators and Bose--Einstein Condensate Dark Matter (BECDM) by Bettoni \& Liberati and collaborators. We first identify SFDM, which is specifically designed to reproduce MOND and is thus constructed in the non-relativistic regime, as the low-acceleration and Newtonian limit of a specific family of Horndeski theories known as RAQUAL. These can then be compared to BECDM, both fully relativistic and generally covariant, by consistently transforming between the Einstein frame and the Jordan frame, and we find the interaction term in the former and the non-minimal coupling term in the latter can be made equivalent up to a linear perturbation. Their kinetic terms are inequivalent. By getting mapped to RAQUAL, SFDM inherits its merits as well as its open issues. In contrast, the jury is still out on BECDM to determine its theoretical and observational standing.

{Motivated by their common condensed-matter inspiration and their shared aim of reconciling MOND-like phenomenology on galactic scales with particle dark matter on larger scales, we investigate the relation between Superfluid Dark Matter (SFDM) and Bose--Einstein Condensate Dark Matter (BECDM). Since SFDM is formulated in the Newtonian regime whereas BECDM is fully relativistic, we first show that the MONDian formulation of SFDM arises as the Newtonian, low-acceleration limit of a Relativistic AQUAdratic Lagrangian (RAQUAL) theory in the Einstein frame. We then transform its covariant interaction sector to the Jordan frame and compare it with BECDM. The phonon--baryon interaction of SFDM maps onto the BECDM derivative coupling to the Einstein tensor, supplemented by a small non-minimal coupling to the Ricci scalar. The interaction sectors are therefore equivalent up to a linear perturbation of the Einstein--Hilbert term. Their kinetic sectors, however, remain inequivalent: the standard quadratic kinetic term of BECDM cannot be mapped onto the non-analytic kinetic term required by SFDM. The two models are consequently related but not dynamically equivalent. This mapping provides a covariant interpretation of the SFDM interaction and clarifies which theoretical properties can be transferred between the two frameworks.}

\end{abstract}

\maketitle

\section{Introduction}
\label{sec:Intro}

% In the zoo of modified gravity models aimed at explaining or replacing dark matter (DM), it would be useful to find some more structure. This is difficult because of the wide range of contexts of applicability of modified-gravity theories: dark matter, dark energy, inflation, quantum gravity. These different contexts have a corpus of literature produced by research communities that do not always talk to each other, leading to different nomenclatures for the same or related theories. An example are k-essence theories from the context of inflation and Relativistic AQUAdratic Lagrangians (RAQUAL) from the context of dark matter modified gravity theories \cite{Bruneton_2007}. The different nomenclatures lead to difficulties in identifying models that are similar or equivalent and limit the cross-pollination of ideas stemming from the different communities and creating more order in the zoo of models.

{The landscape of modified-gravity models developed to explain phenomena commonly attributed to dark matter (DM) would benefit from a clearer organizing framework. Constructing such a framework is challenging because modified-gravity theories arise across widely different domains—including dark matter, dark energy, inflation, and quantum gravity—each with its own literature and research community. Limited communication among these communities has often produced distinct nomenclatures for identical or closely related theories. A notable example is the correspondence between $k$-essence theories, originating in studies of inflation, and Relativistic AQUAdratic Lagrangian (RAQUAL) theories, developed in the context of modified-gravity alternatives to dark matter \cite{Bruneton_2007}. Such terminological fragmentation obscures similarities and equivalences among models, hinders the transfer of insights across fields, and ultimately makes the theoretical landscape more difficult to systematize.

More significantly, however, the difficulty is exasperated by the fact that some models are written in the Einstein frame (EF), while others in the Jordan frame (JF). The concept of EF and JF stems from the equivalence of different representations of scalar–tensor theories of gravity, such as Brans–Dicke theory, based on the freedom to set two of the four free functions in these models arbitrarily through scalar-field redefinitions and conformal transformations of the metric \cite{Flanagan_2004,Sotiriou:2007zu,Kuntz:2026vhs}. In the more general setting of Horndeski theories --- the most general scalar–tensor theories with second-order equations of motion --- it is the broader class of disformal transformations, of which conformal transformations are a special case, that relate different representations of the theory \cite{Bettoni_2013}. In this more general case, there are more free functions in the models, but also more freedom, which in total amounts to a bigger number of physically meaningful frames. 

The possibility of representing the same modified gravity theory in different, physically equivalent\footnote{For precise considerations on what constitutes physical, and also mathematical, equivalence in the sense that we are using it here, the reader is referred to \cite{Flanagan_2004,Kuntz:2026vhs}.} frames increases the difficulty of checking for equivalence between apparently different models. We consider this a fruitful exercise not only for the sake of finding structure within modified-gravity models. More importantly, mapping different theories to each other also facilitates a "transfer" of results, physical insight, and issues between the models. If one theory predicts certain phenomena or suffers from certain theoretical or observational problems, by identifying features between different models, it becomes more evident which results or issues will be shared between models as well. This will help to more quickly discriminate between theoretically and observationally viable models and inviable ones.

Here, we investigate whether such a mapping can be established between Superfluid Dark Matter (SFDM) \cite{Khoury_2015, Khoury_2022, De_Luca_2023} and Non-Minimally Coupled Dark Matter (NMCDM) \cite{Bettoni_2011, Bettoni_2012, Bettoni_2014, Bettoni_2015}. The motivation for comparing these theories lies in their shared aim of combining the successes of Modified Newtonian Dynamics (MOND) on galactic scales \cite{Milgrom_1983, Famaey_2012, Milgrom:2014usa, 1984ApJ...286....7B} with those of particle dark matter on cluster and cosmological scales. Although not every realization of these theories recovers MONDian phenomenology in galaxies, specific formulations do \cite{Khoury_2022, Bruneton_2009}.\footnote{Of course, MOND is not the only framework capable of accounting for observations on galactic scales, and recovering it in the relevant regime is therefore not an indispensable requirement for a modified-gravity theory. Nevertheless, given MOND's notable empirical successes at these scales, agreement with its predictions may be regarded as a promising feature.}

The possibility of a mapping between SFDM and NMCDM is further suggested by their common inspiration in condensed-matter physics. In SFDM, this connection is explicit: the relevant low-energy degrees of freedom are phonons propagating through a Bose--Einstein condensate (BEC). In NMCDM, Bose--Einstein condensation of the scalar field on galactic scales has been proposed as a possible mechanism for dynamically generating its non-minimal coupling to gravity. We refer to NMCDM supplemented by this condensation mechanism as BECDM.

% Besides this informed intuition about a possible connection between the two models, such a link is far from obvious at first glance. First, SFDM is a theory for phonon perturbations in a condensate, whereas BECDM is a theory for the condensate (background) wavefunction. This first obstacle ends up vanishing, however, when one understands that the SFDM action is constructed directly at the level of phonons with an integrated background. The terms in the phonon action are postulated in order to ensure that it reproduces MOND in the relevant regime. It is therefore equivalent to an action of a fundamental scalar field, much like the one for BECDM. {This places on the same theoretical level the phonon field $\pi$ in SFDM and the background scalar field $\phi$ in BECDM, and hence justifies their comparison}. 
%Hence our direct comparison between the two models is justified. 

Despite this physically motivated intuition, a connection between the two models is far from evident at first sight. The first apparent obstacle is that SFDM describes phonon perturbations propagating through a condensate, whereas BECDM describes the background wavefunction of the condensate itself. This distinction becomes less fundamental, however, once one recognizes that the SFDM action is formulated directly at the level of the phonon degrees of freedom. The terms in the phonon action are postulated so as to recover MONDian phenomenology in the relevant regime. In this sense, the resulting theory can be treated as a standard scalar-field theory, much like BECDM. This places the SFDM phonon field $\pi$ and the BECDM background field $\phi$ at the same level of theoretical description, thereby providing a basis for their comparison.

% A less immediate obstacle is the fact that SFDM is constructed in the Newtonian regime from the start, making it not immediately mappable to BECDM, which is fully relativistic. To tackle this difficulty, we identify SFDM as the Newtonian limit of a Relativistic AQUAdratic Lagrangian (RAQUAL) model \cite{Bruneton_2007} written in the Einstein frame with an interaction term that can be directly mapped to a non-minimal coupling term of the scalar field to gravity when written in the Jordan frame. The main computational result of the present work is that, after performing such a mapping, the two end up being equivalent to each other up to a perturbation. In essence, SFDM's interaction term written fully covariantly as a RAQUAL model is a linear perturbation of BECDM's non-minimal coupling term. The kinetic terms, however, are substantially different, the former being non-analytic, and the latter being the standard quadratic one. By getting mapped to RAQUAL, SFDM inherits its merits as well as its open issues. In contrast, the jury is still out on BECDM to determine its theoretical and observational standing.

A subtler obstacle is that SFDM is formulated from the outset in the Newtonian regime, whereas BECDM is a fully relativistic theory, precluding a direct comparison. To overcome this difficulty, we identify SFDM with the Newtonian limit of a Relativistic AQUAdratic Lagrangian (RAQUAL) model \cite{Bruneton_2007}. In the Einstein frame, this relativistic completion contains an interaction term that, upon transformation to the Jordan frame, can be mapped directly onto a non-minimal coupling between the scalar field and gravity.

The principal computational result of this work is that, after performing this mapping, the interaction sectors of the two theories are equivalent up to a perturbative correction. More precisely, the fully covariant SFDM interaction term, expressed within the RAQUAL framework, corresponds to a linear perturbation of the non-minimal coupling term in BECDM. Their kinetic sectors, however, remain substantially different: the SFDM/RAQUAL kinetic term is non-analytic, whereas BECDM possesses the standard quadratic kinetic term. Through its mapping to RAQUAL, SFDM inherits both the strengths and the unresolved issues of that class of theories. By contrast, the theoretical consistency and observational viability of BECDM remain to be fully assessed.

This paper is organized as follows. In Sec.~\ref{Sec:nutshell}, we review the formulations of SFDM and NMCDM relevant to our analysis and introduce BECDM as the realization of NMCDM in which Bose--Einstein condensation dynamically generates the non-minimal coupling. In Sec.~\ref{sec:gencond}, we derive a general condition for obtaining MOND-like baryonic dynamics and establish the relation between an explicit dark-matter--baryon interaction in the Einstein frame and a non-minimal coupling in the Jordan frame. In Sec.~\ref{sec:SFvsBD}, we show that SFDM can be identified with the Newtonian, low-acceleration limit of a RAQUAL theory by matching its gravitational, kinetic, and interaction sectors. In Sec.~\ref{sec:transf}, we transform the relativistic SFDM/RAQUAL theory between frames, compare the resulting non-minimal couplings with those of BECDM, and examine the transformation of the kinetic sector. Finally, in Sec.~\ref{sec:Concl}, we discuss the theoretical and phenomenological implications of the mapping and summarize our conclusions.
}

\section{SFDM and NMCDM in a nutshell}
\label{Sec:nutshell}
In this section we very briefly summarize the main tenets and relevant definitions of the two models we will be comparing. To gain a deeper and more exhaustive understanding, {we refer the reader to} the original papers on Superfluid Dark Matter \cite{De_Luca_2023, Khoury_2015, Khoury_2022}. Phenomenological applications of SFDM can be found in \cite{Berezhiani:2015bqa, Berezhiani:2015pia, Berezhiani:2017tth, Berezhiani:2018oxf, Berezhiani:2019pzd, Berezhiani:2020umi, Berezhiani:2021rjs, Berezhiani:2022buv, Berezhiani:2023vlo, Berezhiani:2025maf}. For studies on the empirical viability of Superfluid Dark matter, see \cite{Hossenfelder_2019, Hossenfelder:2020yko, Mistele:2020qha, Mistele:2022vhh, Mistele:2023wao}. The original literature on Non-minimally Coupled Dark Matter and its variant Bose--Einstein Condensate Dark Matter is \cite{Bettoni_2011, Bettoni_2012, Bettoni_2013, Bettoni_2014, Bettoni_2015}. For interesting phenomenological applications of NMCDM and an empirical test, see also \cite{Benetti_2025, Silveravalle_2025, Gandolfi_2022, Gandolfi:2021jai}.

Throughout this analysis, Superfluid Dark Matter (SFDM) refers specifically to the formulation designed to reproduce MOND within the superfluid cores of galactic halos through a non-analytic phonon action and a phonon--baryon interaction. Observational studies have found this formulation to be less empirically successful than MOND itself \cite{Hossenfelder_2019, Hossenfelder:2020yko, Mistele:2020qha, Mistele:2022vhh, Mistele:2023wao}. More recent developments have instead emphasized scalar-field models with quartic or sextic self-interactions \cite{Berezhiani:2025maf}. Although these models may also form superfluid cores, together with residual superfluid streams extending beyond the tidal-disruption radius, they do not contain the phonon-mediated force required to reproduce MOND within the core. The discussion below therefore concerns the earlier, MOND-reproducing formulation of SFDM whose relation to BECDM is investigated in the subsequent sections.

\subsection{SFDM}
{The low-energy degrees of freedom of a superfluid can be illustrated using the simplest zero-temperature model: a massive complex scalar field with a repulsive quartic interaction,
\begin{equation}
    \mathcal{L}_{\psi}
    =
    -\partial_\mu\psi^{*}\partial^\mu\psi
    -m^2|\psi|^2
    -\frac{g}{2}|\psi|^4\,,
\end{equation}
where $g>0$ ensures that the potential is bounded from below and that the resulting superfluid is stable \cite{Khoury_2022}. This theory is invariant under the global $U(1)$ transformation $\psi\rightarrow e^{i\alpha}\psi$, associated with particle-number conservation. At finite particle density, the homogeneous condensate may be described in the mean-field approximation by
\begin{equation}
    \psi_0(t)=\nu e^{i\mu_{\rm R}t}\,,
    \qquad
    \mu_{\rm R}\simeq m+\mu\,,
\end{equation}
where $\mu_{\rm R}$ and $\mu$ are the relativistic and non-relativistic chemical potentials, respectively. In the non-relativistic limit, the condensate amplitude is related to the particle-number density by $n\simeq 2m\nu^2$.

The condensate spontaneously breaks the global $U(1)$ symmetry, giving rise to a gapless Goldstone mode. Fluctuations around the condensate can be parametrized as
\begin{equation}
    \psi(\boldsymbol{x},t)
    =
    \left[\nu+h(\boldsymbol{x},t)\right]
    e^{i\mu_{\rm R}t+i\pi(\boldsymbol{x},t)}\,.
\end{equation}
Here, $h$ denotes the massive radial mode, while $\pi$ is the gapless phase fluctuation identified with the phonon. At energies and momenta well below the mass of the radial mode, $h$ can be integrated out, leaving an effective theory for $\pi$. For the quartic model above, this procedure gives, at leading order in derivatives,
\begin{equation}
    \mathcal{L}_{\pi}
    =
    \frac{2m^2}{g}X^2\,,
    \qquad
    X
    =
    \mu-m\Phi+\dot{\pi}
    -\frac{(\boldsymbol{\nabla}\pi)^2}{2m}\,,
\end{equation}
where $\Phi$ is the Newtonian gravitational potential. The quartic theory is therefore a useful illustration of how a phonon effective action emerges, but it is not the specific superfluid theory used to reproduce MOND.

More generally, the shift symmetry inherited from the broken $U(1)$ symmetry, together with Galilean invariance, constrains the leading-order zero-temperature phonon Lagrangian to take the form
\begin{equation}
    \mathcal{L}_{\pi}=P(X)\,.
\end{equation}
The function $P$ specifies the equation of state of the superfluid: on the homogeneous background, $P(\mu)$ is the pressure and $n=P_{,X}$ is the particle-number density. The microscopic quartic interaction gives $P(X)\propto X^2$. The MOND-producing formulation of SFDM instead postulates
\begin{equation}
    P_{\rm SFDM}(X)
    =\frac{2\Omega\,(2m)^{3/2}}{3}
    X\sqrt{|X|}\,,
\end{equation}
where $\Omega\sim\sqrt{a_0 M_\textsc{pl}}$ is an energy scale which must be $\mathcal{O}$(meV) so to provide the correct order of magnitude for the MOND acceleration scale $a_0\approx 1.2\times10^{-8}\mbox{cm}/\mbox{s}^2$~\cite{Khoury_2022}. Although this dependence is non-analytic in $X$, it is admissible in an effective theory for collective excitations and corresponds to an equation of state that is analytic in the particle density. In a microscopic description, this scaling can be associated with effective three-body interactions, or equivalently with a sextic scalar potential \cite{Khoury_2022}.

The complete non-relativistic SFDM Lagrangian considered here is therefore
\begin{equation}
\label{eq:sfdmlagr}
    \mathcal{L}_\textsc{sfdm}
    =
    -\frac{1}{8\pi G_N}
    (\boldsymbol{\nabla}\Phi)^2
    +P_\textsc{sfdm}(X)
    +\mathcal{L}_\textsc{b}
    +\mathcal{L}_\textsc{int}\,,
\end{equation}
where $\mathcal{L}_b$ is the baryonic Lagrangian and $\Phi$ is the Newtonian potential. To mediate a MOND-like force, the phonon is coupled directly to the baryonic mass density through
\begin{equation}
    \mathcal{L}_\textsc{int}
    =
    \gamma\frac{\Omega}{M_{\rm Pl}}\,
    \pi(\boldsymbol{x},t)\rho_\textsc{b}\,,
    \label{SFDM-int}
\end{equation}
where $\gamma$ is a dimensionless coupling and $M_{\rm Pl}$ is the reduced Planck mass. This operator explicitly breaks the phonon shift symmetry, but its Planck-suppressed coefficient makes the breaking technically natural from an effective-field-theory perspective. For the appropriate relation among $\gamma$, $\Omega$, and the MOND acceleration scale $a_0$, the resulting phonon-mediated acceleration satisfies~\cite{Khoury_2022}
\begin{equation}
    a_{\pi}\simeq\sqrt{a_0a_\textsc{b}}
\end{equation}
in the static, gradient-dominated regime, where $a_b$ is the Newtonian acceleration generated by baryons. Thus, the non-analytic function $P_{\rm SFDM}(X)$ determines the left-hand side of the MONDian field equation, while the phonon--baryon coupling supplies its baryonic source.

The comparison with BECDM is performed at the level of these effective scalar actions. In particular, the condensate wavefunction is not identified with the phonon perturbation. Rather, once the massive radial mode has been integrated out, $\pi$ is the sole low-energy scalar degree of freedom and its effective action can be placed on the same formal footing as the BECDM scalar-field action. The mapping developed below provides a possible covariant interpretation of the phonon--baryon interaction, which is introduced phenomenologically in SFDM to recover the MOND force law.

Finally, the phonon-mediated force exists only where the dark-matter particles have condensed and thermalized into a coherent superfluid. In galactic halos this is expected to occur within a central superfluid core, while the outer halo remains in a normal, approximately collisionless phase. Within the core, the total acceleration acting on baryons is
\begin{equation}
    \boldsymbol{a}
    =
    \boldsymbol{a}_\textsc{b}
    +\boldsymbol{a}_\textsc{dm}
    +\boldsymbol{a}_{\pi}\,,
\end{equation}
where $\boldsymbol{a}_{\rm DM}$ is the ordinary gravitational acceleration generated by the dark-matter density. The MOND limit is recovered only when the phonon contribution dominates over $\boldsymbol{a}_{\rm DM}$. Since this condition is restricted to the superfluid core, SFDM does not reproduce MOND throughout the entire halo \cite{Berezhiani:2025maf, De_Luca_2023}; this restricted domain is largely responsible for its different observational performance relative to MOND \cite{Hossenfelder_2019, Hossenfelder:2020yko, Mistele:2020qha, Mistele:2022vhh, Mistele:2023wao}.
}

\subsection{NMCDM}
{Non-Minimally Coupled Dark Matter is a relativistic scalar-tensor theory of gravity with the metric $\tilde{g}_{\mu \nu}$ and the scalar field $\phi$ being the gravitational propagating degrees of freedom. In addition, we shall collectively label as $\psi$ the barionic matter degrees of freedom.}
Schematically, the action is composed of the standard Einstein--Hilbert term $S_\textsc{eh}$\footnote{Following the authors of Refs. \cite{Bettoni_2011, Bettoni_2012}, we will consistently neglect the tiny cosmological constant term in this work.}, the action for baryons $S_\textsc{B}$, for the scalar field which plays the role of DM $S_\textsc{dm}$, and a non-minimal coupling term $S_\textsc{nmc}$
\begin{equation}
\label{eq:nmcdmaction}
    S = S_\textsc{eh}[\tilde{g}_{\mu \nu}] + S_\textsc{B}[\tilde{g}_{\mu \nu},\psi] + S_\textsc{dm}[\tilde{g}_{\mu \nu}, \phi] + S_\textsc{nmc}[\tilde{g}_{\mu \nu}, \phi]
\end{equation}
The idea of the authors of this framework is to unite the successes of MOND on galactic scales and of particle DM on cluster and cosmological scales. The NMC term can lead to an effective modification of the way baryons react to gravity and thus cause a MOND-like universal modification to baryon dynamics. We will formalize this statement in the next section by talking about the equivalence of the Jordan and the Einstein frame, but it can intuitively be understood as follows. If DM couples non-minimally to curvature, and baryons couple in a standard way to gravity, by associativity we will inevitably obtain an indirect influence of the non-minimal DM coupling to the baryon dynamics.

The inclusion of a non-minimal coupling term might seem ad-hoc at first, but it is actually quite natural. Its absence is generally a choice made in the name of simplicity and minimality. If it becomes useful to explain observations (indeed, there is evidence that it could very well match observations by resolving the cuspiness of galactic DM halo cores \cite{Bettoni_2014, Gandolfi_2022, Gandolfi:2021jai}), there is no reason not to consider it. The presence of a non-minimal coupling is allowed by symmetry, thus it would naturally appear in an effective field-theory framework, and for a fluid, which is a common approximation for the field representing any cosmological ingredient of the Universe, it is even expected. Indeed, a non-minimal coupling term expresses coupling to spacetime curvature (i.e., a coupling to second derivatives of the metric \cite{Bettoni_2014}). Such a coupling felt over a region of spacetime as opposed to point-like is expected for fluids, these being systems composed of elements no smaller than the mean free path of the underlying particles. In essence, the fluid description erases sensibility on scales smaller than this length scale.

The question is not how to argue for the presence of the non-minimal coupling in general, but rather how to activate it \emph{only} on galactic scales, and maintain it turned off on larger scales. The fluid description of DM could prove useful here again: indeed, DM is expected to be a very weakly self-interacting field, meaning that, in a particle-description, the mean free path of the particles will be quite large and could be of the order of the curvature radius within a galaxy \cite{Gandolfi_2022, Gandolfi:2021jai}. This could also be the reason why this non-minimal coupling is relevant for DM but not for baryons, although the equivalence principle would require it to be valid for both. Baryons have much smaller mean free paths, these being much more strongly interacting, and the discrepancy between this length scale and the curvature radius on all relevant scale means the non-minimal coupling term for them would be negligible \cite{Bettoni_2011}. Formalizing and quantifying these statements within the framework of effective field theory of hydrodynamics is a work in progress.

There could be an even more natural way for the NMC to arise dynamically only for DM on the relevant scales. If the necessary conditions are present, the DM scalar field can undergo Bose--Einstein condensation, developing a macroscopic coherence length {(corresponding to the so called the healing length, in standard BEC), below which all sensitivity to perturbations is washed out \cite{Barcel__2005}. If such a coherence length is of the order of the curvature radius it is then natural to conceive it might lead to non-minimal couplings.  

Furthermore, one might reasonably conjecture that such condensation may happen precisely at galactic scales because the increased DM density there, can drive an increase of the critical temperature and thus a more easily surpassable threshold for condensation \cite{Bettoni_2014}. Furthermore, the critical temperature for condensation depends on properties of the trapping potential the system is immersed in. It important to stress that so far this is no more than a plausible scenario as we are still missing a detailed discussion of Bose--Einstein condensation in a "Newtonian trap" (i.e.~a confinement provided via a Newtonian gravitational potential). So it remains to be clarified if such a mechanism would closely mimic what is well known for harmonic traps and under which conditions it might become relevant exactly at galactic scales. In this sense, BECDM is merely NMCDM with a particular mechanism for the appearance of the non-minimal coupling.

In the NMCDM and BECDM literature, non-minimal coupling terms are proposed based on the criteria of keeping the scalar field equations second-order in derivatives and on the introduction of a new length scale through dimensional analysis. For example, the term $\phi R$ is not considered in \cite{Bettoni_2014}, precisely because it does not introduce a new length scale, apart from not respecting scalar field shift symmetry. The NMC terms in the field-theoretic framework proposed in the BECDM paper \cite{Bettoni_2014} fulfilling the above criteria are
\begin{equation}
\label{eq:nmcterms}
    XR \qquad {\rm and} \qquad G^{\mu \nu}\nabla_{\mu} \phi \nabla_{\nu} \phi 
\end{equation}
where $X=-g^{\mu \nu}\partial_\mu \phi \partial_\nu \phi /2$ is the standard kinetic term of a scalar field, $R$ is the Ricci scalar, and $G^{\mu \nu}$ is the Einstein tensor. However, these terms are equivalent to each other up to a total derivative and higher-order perturbations in field derivatives that are neglected\footnote{This is a good approximation both in the cosmological context and in that of static dark matter halos.}
\begin{equation}
\label{eq:nmctermsequiv}
    XR + (\Box\phi)^2 - (\nabla_{\mu}\nabla_{\nu}\phi)^2 = G^{\mu \nu} \nabla_\mu \phi \nabla_\nu \phi + \mbox{tot. der.}
\end{equation}
Therefore, {one is free to choose to work with the latter only}. 

Dimensional requirements imply that such term must be charactrized by a a prefactor of dimensions length squared which in the simplest scenario can be taken as a constant $L^2$ so that, in conclusion, the NMC term adopted in NMCDM/BECDM takes the form
\begin{equation}
\label{eq:becnmcterm}
  S_\textsc{nmc}\propto  - L^2\int d^4x\sqrt{\tilde{g}} \, \tilde{G}^{\mu \nu}\tilde{\nabla}_{\mu} \phi \tilde{\nabla}_{\nu} \phi\,,
\end{equation}
{the latter being added to the usual total Lagrangian describing a standard $\Lambda$CDM scenario.} We note that the minus sign has been chosen due to the viability of the resulting phenomenological consequences for the cores of dark matter halos and of ultracompact objects \cite{Benetti_2025, Silveravalle_2025}.

A final remark: since our sought comparison with SFDM requires to work within a field-theoretic language, we will not discuss here the formulation of NMCD in hydrodynamic language\footnote{In the hydrodynamic framework, the main quantities one works with are the four-velocity and density of a fluid. In contrast, in the field-theoretic formalism, one works directly with the field and its derivatives.}. Still, let us only recall that a mapping between the two formulations, based on expressing the fluid quantities through the field and its field derivatives, can be performed \cite{Bettoni_2012}.

\section{A general modified dynamics for baryons}
\label{sec:gencond}
In this Section, we elaborate how a general modified dynamics for baryons can be obtained on the level of the action. To do so, our starting point will be $\Lambda$CDM scenario supplemented by a very weak DM-baryons coupling. The action, written in this way, is in the Einstein frame. Upon transformation to the Jordan frame, we will see that a non-minimal coupling emerges which already bears resemblance to $S_\textsc{nmc}$ described by Eq.~\eqref{eq:becnmcterm} within the NMCDM framework's action, Eq.~\eqref{eq:nmcdmaction}. This will be the basis for comparison between SFDM, written in the Einstein frame, and NMCDM, written in the Jordan frame

\subsection{Interacting DM in an Einstein frame}
Let us start by writing down a general theory of gravity with baryons and dark matter but postulate that the latter have also a (very weak) interaction term $S_\textsc{int}$. One can then write
\begin{equation}
    S = S_\textsc{eh}[g_{\mu \nu}] + S_\textsc{b}[g_{\mu \nu},\psi] + S_\textsc{dm}[g_{\mu \nu}, \phi] + S_\textsc{int}[g_{\mu \nu}, \psi, \phi] \qquad \text{(EF-int)}\,,\label{eq:efint}
\end{equation}
%with the Einstein-Hilbert action, a baryonic and dark matter action and the interaction thereof. 
recalling that $\psi$ and $\phi$ are collective fields denoting baryons and DM, respectively, whereas $g_{\mu \nu}$ is the standard gravitational metric \cite{Bettoni_2011}. We dub the action written in this way "EF-int", standing for the variant of the Einstein frame with an explicit interaction term between baryons and dark matter. 

We can now ask what the interaction term should look like to provide an effective metric affecting only the baryons, as this, in a suitable Newtonian limit, could mimic a MONDian behaviour. Hence we shall explicitly ask that
\begin{equation}
    S_\textsc{b}[\psi, g_{\mu \nu}] + S_\textsc{int}[\psi, \phi, g_{\mu \nu}] \approx  S_\textsc{b}[\psi, \tilde{g}_{\mu \nu} = g_{\mu \nu} + h_{\mu \nu}] \qquad \text{(EF-mond)}\label{eq:efmond} \, ,
\end{equation}
where we require for the difference $h_{\mu \nu}$ between the physical $\tilde{g}_{\mu \nu}$ and the gravitational metric $g_{\mu \nu}$ to be small as a consequence of the smallness of the DM-baryons interaction. 

Since this is all still in the Einstein frame ($S_\textsc{eh}$ and $S_\textsc{dm}$ are left untouched, without any non-minimal coupling), we have dubbed this form of the action with modified dynamics for baryons "EF-mond". The possibility of writing the action in this way is the general necessary condition for a theory to reproduce a modified dynamics for baryons. The reason is that, in this case, baryons do not follow the geodesics of the gravitational metric, but the ones of the effective, so-called "physical", metric $\tilde{g}_{\mu \nu}$. Whether this modified dynamics will be precisely the MOND one \cite{Milgrom_1983, Milgrom:2014usa} depends upon further conditions, of course. 

\subsection{Disformal transformations in two-geometry descriptions of physics}

This form of the action expresses a physical insight which was thoroughly analyzed by Bekenstein, namely that gravitational and physical geometry need not be the same. It just so happens that they are in GR, making this theory \emph{a one-geometry description of physics}~\cite{Bekenstein_1993}, but this does not mean that they have to be in every theory of gravity. More general theories would have a separate gravitational geometry described by $g_{\mu \nu}$, describing the curvature of spacetime, and a physical geometry $\tilde{g}_{\mu \nu}$, in which matter plays out its dynamics \cite{Bekenstein_1993}. 

These theories could then be called \emph{two-geometries approaches}. They would violate the strong equivalence principle but preserve weak equivalence, and one of the defining features of the theory would be the prescription of the relation between these two geometries. While most known theories assume the relation is a conformal transformation, it was showed in~\cite{Bekenstein_1993} that a more general transformation is also possible, i.e.~a disformal one. It bears this name because it is a change of local units of length with a preferred direction along the gradient of the scalar field in the theory. Explicitly, 
\begin{equation}
\label{eq:disftr}
    \tilde{g}_{\mu \nu} = A(\phi, X)g_{\mu \nu} + B(\phi, X)\nabla_\mu \phi \nabla_\nu \phi \, ,
\end{equation}
where the coefficients $A$ and $B$ are for now unspecified functions of the scalar field and its kinetic term.

It was later found, as was mentioned, that these more general transformations, if invertible, maintain the Horndeski action covariant \cite{Bettoni_2013, Zumalac_rregui_2013}, amounting to a renormalization of the Horndeski functions $G_i$. To do this in a way that keeps the differential equations second-order, the dependence on $X$ of the coefficients $A$ must be abandoned \emph{if} the function $G_4$ appearing in the action term $G_4R$ with the Ricci scalar is a function of $X$ as well as of the scalar field: $G_4(\phi, X)$~\cite{Bettoni_2013, Zumalac_rregui_2013}. Since one of the possible NMC terms we want to consider is $XR$, as shown in Eq.~\eqref{eq:nmcterms}, we will set $A(\phi, X) = A(\phi)$, but doing the same for $B$ is not necessary, as is done in the above Refs. We will remain general and retain $B=B(\phi, X)$\footnote{In fact, there are arguments that at least one of the functions $A$ and $B$ must be a function of $X$ as well as the field in order to retain a Lorentzian signature for the physical metric, although there might be ways to ensure the correct signature differently. For more details, we refer to \cite{Bettoni_2013, Bruneton_2007} and references therein.}, although this will not be of further relevance here.

\subsection{DM-baryon interaction for modified baryon dynamics}

Given the above discussion, we shall further restrict the (small) difference between the physical and gravitational metric $h_{\mu \nu} \equiv \tilde{g}_{\mu \nu} - g_{\mu \nu}$ to be of the disformal form \footnote{Let us comment that, for $h_{\mu \nu}$ to be consistently a small quantity, $A(\phi)$ must be close to unity, and $B(\phi, X)$ must be close to zero and/or field derivatives must be small. The smallness of field derivatives was already required for (\ref{eq:nmctermsequiv}) and should be valid for static DM halos. Of course, our framework simply loses its validity in the cases where the combination $B\nabla_\mu \phi \nabla_\nu \phi$ is not small.}
\begin{equation}
\label{eq:formofh}
   h_{\mu \nu}(\phi, X) = (A(\phi)-1)g_{\mu \nu} + B(\phi, X)\partial_\mu \phi \partial_\nu \phi
\end{equation}

Whenever this quantity is sufficiently small we can expand the r.h.s.~of Eq.~\eqref{eq:efmond} so to get, at first order in $h_{\mu\nu}$
\begin{equation}
    S_\textsc{b}[\psi, \tilde{g}_{\mu \nu} = g_{\mu \nu} + h_{\mu \nu}]\approx S_\textsc{b}[\psi, g_{\mu \nu}]-\frac{1}{2}\int d^4x\sqrt{-g}T_\textsc{b}^{\mu \nu}h_{\mu \nu}(\phi, X) \, ,
\end{equation}
where we have used $T_\textsc{b}^{\mu \nu} = -\frac{2}{\sqrt{-g}}\frac{\delta S\textsc{b}}{\delta g_{\mu \nu}}$ and $\delta g_{\mu \nu} = h_{\mu \nu}$. We have thus shown that, to satisfy the condition for obtaining a modified dynamics for baryons as in Eq.~\eqref{eq:efmond} for small $h_{\mu \nu}$, the interaction term appearing in the action~\eqref{eq:efint} must be of the form

\begin{equation}
\label{eq:neccesint}
    S_\textsc{int}[\psi, \phi, g_{\mu \nu}] = -\frac{1}{2}\int d^4x\sqrt{-g}T_\textsc{b}^{\mu \nu}(\psi)h_{\mu \nu}(\phi, X) \, .
\end{equation}

This equivalent to say that in this limit the action~\eqref{eq:efint} takes the form of standard GR plus a minimally coupled scalar field, minimally coupled baryons, \emph{and} a particular interaction term between baryons $\psi$ and DM $\phi$. This is the form allowing for an effective geometrical interpretation of the DM interaction with baryons, and as such it is a necessary (but not sufficient) condition for reproducing a MOND-like behavior in this setting~\cite{Bettoni_2011, Bettoni_2012}. 
%Note, that all the fields have to be included in the above stress energy tensor, consequently $S_\textsc{int}$ implies also a (very weak) self-interaction of the DM field with itself.

\subsection{Jordan frame}
We can now transform the entire action~\eqref{eq:efint} with the above interaction term~\eqref{eq:neccesint} into the Jordan frame (JF) via a disformal transformation~\eqref{eq:disftr}, i.e.~writing everything in terms of the physical metric. We obtain an action of the form (\ref{eq:nmcdmaction}) which we reiterate here \cite{Bettoni_2011, Bettoni_2012, Bruneton_2009}
\begin{equation}
\label{eq:jfact}
    S = S_\textsc{eh}[\tilde{g}_{\mu \nu}] + S_\textsc{b}[\tilde{g}_{\mu \nu},\psi] + S_\textsc{dm}[\tilde{g}_{\mu \nu}, \phi] + S_\text{nmc}[\tilde{g}_{\mu \nu}, \phi] \qquad \text{(JF)}
\end{equation}
where
\begin{equation}
\label{eq:nmctermjf}
    S_\text{nmc}[\tilde{g}_{\mu \nu}, \psi, \phi] = -\frac{1}{16\pi G_N}\int d^4x \sqrt{\tilde{g}}\,G^{\mu \nu}(\tilde{g}_{\mu \nu})h_{\mu \nu}(\phi, X(\tilde{g}_{\mu \nu}))
\end{equation}
where $G^{\mu \nu}(\tilde{g}_{\mu \nu})$ is the Einstein tensor written with the physical metric. A cosmological constant term has been consistently neglected\footnote{It is of the form $\Lambda \tilde{g}^{\mu \nu} h_{\mu \nu}$, where $\Lambda$ is the cosmological constant measured to be extremely small. Given that $h_{\mu \nu}$ is also a small quantity, this term can, indeed, be safely neglected.}. We have also made explicit that the kinetic term needs to be computed with the physical metric $X(\tilde{g}_{\mu \nu})$, as is consistent in this frame. Whether this non-minimal coupling term $S_\text{nmc}$ matches the one in NMCDM/BECDM $S_\textsc{nmc}$, given in Eq.~\eqref{eq:becnmcterm}, depends on the form of $h_{\mu \nu}$ as we shall see.

As a summary, in the Jordan frame:
\begin{itemize} 
\item A non-minimal coupling between DM and the metric appears;
\item Gravity is not necessarily described by a purely Einstein-Hilbert action due to the non-minimal coupling;
\item The action is written with the physical metric $\tilde{g}_{\mu \nu}$.
\end{itemize}
It becomes clear that the NMCDM model~\eqref{eq:nmcdmaction} is written in the Jordan frame, with a non-minimal coupling of the form~\eqref{eq:nmctermjf}, that can be seen as emerging from a corresponding interaction term~\eqref{eq:neccesint} in the Einstein frame. To summarize, in the Einstein frame:
\begin{itemize}
\item An interaction term between DM and baryons appears as well as a self-interaction term for DM;
\item Gravity is described by a purely Einstein-Hilbert action;
\item The action is written with the gravitational metric $g_{\mu \nu}$.
\end{itemize}
SFDM, being a model that aims to reproduce MOND in superfluid galactic cores, is written in the Einstein frame, specifically in its form EF-mond (\ref{eq:efmond}). To draw any link between SFDM and NMCDM, we will have to use the above transformation between frames. Before doing this, however, we recognize that SFDM is a Newtonian theory, whereas NMCDM is fully relativistic. We proceed by looking for existing relativistic theories, of which SFDM could be the Newtonian limit.
}

\section{SFDM as RAQUAL}
\label{sec:SFvsBD}

\subsection{RAQUAL action and kinetic term}

RAQUAL stands for "Relativistic AQUAdratic Lagrangians" and it is the name for sub-families of Horndeski models with a non-quadratic kinetic term and non-coinciding gravitational and physical metric. Therefore, {such class of theories can be seen as strictly related to the so-called ``k-essence'' Lagrangians, supplemented by the additional assumption of a two-geometries description.} In particular, the RAQUAL action takes the form
\begin{equation}
\label{eq:raqualact}
\begin{split}
    S = \frac{1}{4\pi G_N} \int d^4x \sqrt{-g}\left\{\frac{R}{4}-\frac{1}{2}f(\phi, X) - V(\phi) \right\} + S_\textsc{b}[\psi, \tilde{g}_{\mu \nu} ]\,,
\end{split}
\end{equation}
where for conciseness we have dropped the cosmological constant term. Similarly, the potential $V(\phi)$ was written for completeness but will be henceforth unnecessary and therefore set to zero. 

{The connection between such relativistic actions and MOND is found by realizing that a RAQUAL kinetic term $f(\phi, X) = f(X) \to l_0X^{3/2}$ --- where the constant $l_0$ has the dimension of length --- reproduces MOND in the Newtonian limit for small $X>0$, see e.g.~\cite{Bruneton_2007}\footnote{To be precise, these authors use a function $f(s)$ where $s$ is defined with a positive sign as $s=g^{\mu \nu}\partial_\mu \phi \partial_\nu \phi$, and they quote only a result for the derivative $f'(s) \propto \sqrt{s}$.}. This limit in turn corresponds to small accelerations $a<a_0$ with $a_0$ being the limiting acceleration below which MONDian behavior starts to manifest.  }

The above aquadratic kinetic term is therefore precisely the kinetic term imposed in Superfluid Dark Matter~\cite{Khoury_2022, Berezhiani:2025maf}.\footnote{For the sake of precision let us stress that in ~\cite{Khoury_2022, Berezhiani:2025maf} it explicitly remarked that $X^{3/2}$ is short-hand for $X\sqrt{|X|}$, in case $X$, which has anyway a sign-definition ambiguity in the extant literature, ends up being negative.}. The next step is to look for a correspondence between RAQUAL's baryonic action term and SFDM's interaction term. 

In the RAQUAL model, it is posited that

\begin{equation}
\label{eq:raqualconf}
    \tilde{g}_{\mu \nu} = e^{2\alpha \phi}g_{\mu \nu} \, ,
\end{equation}
i.e.~a \emph{conformal} transformation relates the physical and gravitational metric, with $A(\phi)=e^{2\alpha \phi}$ where $\alpha$ is a small constant constrained by Cassini spacecraft observations to be $\alpha^2<10^{-5}$ \cite{Bruneton_2007}. This assumption is required to obtain the correct r.h.s. of the equations of motion (EoM). 

On the other hand, the recovery of the r.h.s. of the Poisson-like equation associated to MOND in SFDM, requires the introduction of an interaction Lagrangian of the form Eq.~\eqref{SFDM-int}. We can rescale the phonon field $\pi$ to include the small constant $\frac{\Omega}{M_{pl}}\sim10^{-31}$ into the field definition $\phi \equiv \frac{\Omega}{M_{pl}} \pi$. With this, the DM-baryon interaction term postulated in SFDM becomes

\begin{equation}
        \mathcal{L}_\textsc{int}
  \propto 
    \phi(\boldsymbol{x},t)\rho_\textsc{b}\,,
\end{equation}
where the proportionality indicates that we can still redefine the field $\phi$ by absorbing dimensionless constants in and out of it. We shall now show that the RAQUAL conformal transformation assumption and the SFDM interaction term are equivalent.

\subsection{RAQUAL-SFDM equivalence}
The above discussion shows that this equivalence is not immediately evident simply because the RAQUAL baryon term is written already in EF-mond Eq.~\eqref{eq:efmond}, whereas SFDM, in assuming the presence of an interaction term, positions itself within EF-int Eq.~\eqref{eq:efint}, in addition to being explicitly non-relativistic. Let us then start from the interaction term Eq.~\eqref{eq:neccesint} that we established is necessary to get a modified dynamics for baryons. Inserting $h_{\mu \nu}$ as in Eq.~\eqref{eq:formofh} into this interaction term, we obtain (for brevity, legibility, and consistency with SFDM notation, we work with the Lagrangian terms as opposed to the action)
\begin{equation}
    \mathcal{L}_\textsc{int}=-\frac{1}{2}(A-1)T_\textsc{b} - \frac{B}{2}T_\textsc{b}^{\mu \nu}\nabla_{\mu}\phi\nabla_{\nu}\phi
\end{equation}
where $T_\textsc{b} = T_\textsc{b}^{\mu \nu}g_{\mu \nu}$ is the trace of the baryon stress-energy tensor (SET). Since we are interested in halos, which can be well approximated as stationary spherically symmetric DM configurations, we will assume $\phi = \phi(r)$, thus recognizing that $T_\textsc{b}^{\mu \nu}\nabla_{\mu}\phi\nabla_{\nu}\phi = (\partial_r\phi)^2T_\textsc{b}^{rr}$. Substituting fluid quantities for the baryonic SET
\begin{equation}
    \mathcal{L}_\textsc{int}=-\frac{1}{2}(A-1)(3p_\textsc{b} - \rho_\textsc{b} c^2) - \frac{B}{2}(\partial_r\phi)^2p_\textsc{b} g^{rr}
\end{equation}
where we have restored a factor of $c^2$ that had previously been set to unity. Now we can clearly see that, in the Newtonian limit $c\to\infty$, $\rho_\textsc{b}c^2 \gg p_\textsc{b}$. Thus, the interaction Lagrangian simplifies to

\begin{equation}
    \mathcal{L}_\textsc{int}\approx \frac{1}{2}(A-1)\rho_\textsc{b}c^2
\end{equation}
which has the exact form required in SFDM if we choose $A(\phi)=1+\phi$. We conclude that we can recover the EF-int form \eqref{eq:efint} of the SFDM action from the RAQUAL general relativistic theory written in EF-mond \eqref{eq:efmond} form.

If we now recall that in this framework, $h_{\mu \nu}$ is a small perturbation, we recognize that $A$ must be close to unity, whence the field $\phi$ must also be a small perturbation. Indeed, recall that $\phi(\mathbf{x}, t) = \frac{\Omega}{M_{pl}}\pi(\mathbf{x}, t)$ (dimensionless) was a redefinition of $\pi(\mathbf{x}, t)$ to absorb into the field the tiny constant $\frac{\Omega}{M_{pl}}\sim 10^{-31}$. The phonon field $\pi(\mathbf{x}, t)$ itself was already a small quantity, being by definition a phase perturbation of the condensate wavefunction. Therefore we can complete $A(\phi) = 1+\phi\approx e^\phi$. By means of field-redefinition, we can always extract the smallness of this field in the form of a constant factor in front of it $\phi \to2\alpha\phi$. With this, we recover the precise form of the assumed physical metric in RAQUAL, Eq.~\eqref{eq:raqualconf}.

For completeness, let us make it explicit that the Einstein--Hilbert terms of RAQUAL and SFDM are also the same (the two theories are both formulated in an Einstein frame). So, to conclude this section, we can restate that the Einstein--Hilbert, the kinetic, and the interaction/baryonic action terms between RAQUAL and SFDM are equivalent, noting that in the latter, they are written in EF-mond form, in the Newtonian, low-acceleration regime.

\section{SFDM vs NMCDM}
\label{sec:transf}
\subsection{From the interaction term to the NMC term}
Let us briefly recapitulate. We have found the difference between physical and gravitational metric to be
\begin{equation}
\label{eq:hforsfdm}
    h_{\mu \nu}(\phi, X) = \phi g_{\mu \nu} +B(\phi, X)\nabla_\mu \phi \nabla_\nu \phi \approx (e^\phi-1)g_{\mu \nu} + B(\phi, X)\nabla_\mu \phi \nabla_\nu \phi
\end{equation}
which reduced the standard RAQUAL action Eq.~\eqref{eq:raqualact} to the SFDM one \eqref{eq:sfdmlagr} in the corresponding non-relativistic limit, both of which are written in the Einstein frame. Note that the disformal term with coefficient $B(\phi, X)$ dropped out of the interaction term in this limit. 

A necessary condition for RAQUAL/SFDM to yield a MOND-like behavior is that the interaction term be of the form~\eqref{eq:neccesint}. Transforming the Einstein frame action \eqref{eq:efint} with this interaction term into the Jordan frame~\eqref{eq:jfact} yields the NMC term \eqref{eq:nmctermjf}, into which we can now insert the $h_{\mu \nu}$ from Eq.~\eqref{eq:hforsfdm}, so as to check for compatibility between RAQUAL/SFDM, and NMCDM. 

We obtain
\begin{equation}
    S_\textsc{nmc}[\tilde{g}_{\mu \nu}, \psi, \phi] = -\frac{1}{16\pi G_N}\int d^4x \sqrt{\tilde{g}}\,G^{\mu \nu}(\tilde{g}_{\mu \nu})[\phi (\tilde{g}_{\mu \nu} - h_{\mu \nu}) + B(\phi, X)\nabla_\mu \phi \nabla_\nu \phi]
\end{equation}
where, we have inserted $g_{\mu \nu} = \tilde{g}_{\mu \nu} - h_{\mu \nu}$ in order to write everything in terms of the physical metric as is proper in the Jordan frame. Now, one could substitute Eq.~\eqref{eq:hforsfdm} into $h_{\mu \nu}$ again and iterate the above process, but this would increase the order in $h_{\mu \nu}$ that is being considered. To remain consistently at first order in $h_{\mu \nu}$, this term must simply be neglected without iteration. Using the trace of the Einstein tensor $G^{\mu \nu}(\tilde{g}_{\mu \nu})\tilde{g}_{\mu \nu} = -R(\tilde{g}_{\mu \nu})$, we finally obtain the two NMC terms

\begin{equation}
    S_\textsc{nmc}[\tilde{g}_{\mu \nu}, \psi, \phi] = \frac{1}{16\pi G_N}\int d^4x \sqrt{\tilde{g}}[\phi \tilde{R}- B(\phi, X)\tilde{G}^{\mu \nu}\nabla_\mu \phi \nabla_\nu \phi]
\end{equation}
where for brevity we write $\tilde{R} \equiv R(\tilde{g}_{\mu \nu})$ and $\tilde{G} \equiv G(\tilde{g}_{\mu \nu})$. 

The first term can be added to the Einstein--Hilbert action resulting in an NMC of the scalar field to the Ricci scalar through $A(\phi) = 1+\phi$, while the second term remains as an NMC of the field gradient to the Einstein tensor so that
\begin{equation}
\label{eq:finalehandnmc}
    S_\textsc{eh}[\tilde{g}_{\mu \nu}] +  S_\textsc{nmc}[\tilde{g}_{\mu \nu}, \phi] = \frac{1}{16\pi G_N}\int d^4x \sqrt{\tilde{g}}\,[A(\phi) \tilde{R}- B(\phi, X)\tilde{G}^{\mu \nu}\nabla_\mu \phi \nabla_\nu \phi]
\end{equation}
The second term is exactly of the BECDM form (\ref{eq:becnmcterm}) if 
\begin{equation}
    \frac{B(\phi, X)}{16\pi G_N} \equiv L^2
    \label{eq:BECmapCond}
\end{equation}
i.e.~if $\frac{B(\phi, X)}{16\pi G_N}$ is chosen to be a constant with dimensions length squared \cite{Bettoni_2014}. 

So far, we had assumed $B$ to free to be chosen at will, as far as identifying SFDM as a Newtonian, low-acceleration limit of RAQUAL goes. However, there is a caveat from the theoretical point of view, because $A$ and $B$ need to satisfy constraints in order for $\tilde{g}_{\mu \nu}$ to still be a valid metric. In particular, it must have Lorentzian signature -- otherwise, the matter equations of motion are elliptic and the Cauchy problem is not well-posed. Indeed, to ensure a Lorentzian physical metric, one obtains the following constraint \cite{Bettoni_2013, Bruneton_2007, PhysRevD.111.024053}
\begin{equation}
    A(\phi) - 2XB(\phi, X)>0
    \label{eq:condLore}
\end{equation}
Note that this is a necessary but not sufficient condition for hyperbolicity, as is analyzed extensively in \cite{PhysRevD.111.024053}. 

Now, approximating $A\approx 1$, and imposing the BECDM correspondence condition \eqref{eq:BECmapCond} we find $X\lesssim(32 \pi G_N L^2)^{-1}$. Let us now recall that  in the absence of large gradients in the DM field, $L^2$ must be of the order of the curvature radius squared $\sim |R|^{-1/2}$ for the NMC term (\ref{eq:nmctermsequiv}) to be relevant. Assuming that the stress-energy of the spacetime in the region of interest is dominated by pressureless DM with density $\rho_{DM}$, we can write $|R|^{-1} = (8\pi G_N \rho_{DM})^{-1}$. At the same time, $-X = \frac{1}{2}(\partial_r\phi(r))^2 = \rho_{DM}$ when there is no scalar field potential, so the condition Eq.~\eqref{eq:condLore} becomes $-1\lesssim1/4$, i.e.~it is satisfied. 

This order-of-magnitude estimate yields a satisfactory result, meaning that imposing Eq.~\eqref{eq:BECmapCond} does not cause problems for the well-posedness of the Cauchy problem of matter when deriving BECDM in the Jordan frame from the action in the Einstein frame necessary to have an effective modified dynamics for baryons. It also maintains the smallness of $h_{\mu \nu}$, since $16 \pi G_N \sim L_p^2$ in natural units, where $L_p$ is the incredibly small Planck length.

To conclude, the slight disagreement of RAQUAL/SFDM with BECDM becomes apparent from the first term on the r.h.s. in eq. (\ref{eq:finalehandnmc}). In BECDM, the Einstein--Hilbert term is unmodified and there is no non-minimal coupling of the scalar field directly to the Ricci scalar. In~\cite{Bettoni_2014} it is explicitly mentioned  that this term is viable in a Horndeski framework, albeit it would entail a shift-symmetry breaking. Ultimately, it is normally neglected in the NMCDM/BECDM literature  because it does not introduce a new length scale. 

Furthermore, one could also make the argument that $A(\phi) = 1+\phi$ is very close to unity, since we established that, in our context of spherically symmetric DM halos, the scalar field is a very small radial perturbation $\phi(r)\ll1$. In this sense, in SFDM's relativistic formulation, the interaction term ends up being equivalent up to a small radial perturbation of BECDM's non-minimal coupling term, \emph{if} $\frac{B(\phi, X)}{16\pi G_N} = L^2$.

\subsection{Transformation of the kinetic term}
BECDM uses a standard kinetic term quadratic in field derivatives in the DM action \cite{Bettoni_2014} in the Jordan frame, not an aquadratic one. The question is whether, once one transforms also the kinetic term from the Jordan frame to the Einstein frame one could recover an aquadratic term. 

We already know that this is not the case \cite{Bettoni_2011, Bettoni_2012}, i.e., the kinetic term does not change its structure upon this transformation, \emph{if} $h_{\mu \nu}$ is considered a small perturbation, which it must be so to derive the crucial interaction term \eqref{eq:neccesint}. From this expression we conclude that this amounts to considering $B\nabla_{\mu}\phi\nabla_{\nu}\phi$ and $[A(\phi)-1]$ to be small quantities, in which we have been consistent\footnote{This is also the reason why the terms that are second order in two field derivatives in expression \eqref{eq:nmctermsequiv} have been neglected.}. 

In fact, Eqs. \eqref{eq:efint} and \eqref{eq:jfact} are to be taken literally, in the sense that $S_\textsc{eh}$, $S_\textsc{b}$, and $S_\textsc{dm}$ are, in both of these equations, the \emph{same} functional, but of a different metric in the two cases. Therefore, it is expressed already in the equations, which we merely reported here from Refs. \cite{Bettoni_2011, Bettoni_2012}, that the DM kinetic term cannot change structure. We have also explicitly checked this by performing the disformal transformation Eq. \eqref{eq:disftr} between action \eqref{eq:efint} and \eqref{eq:jfact}\footnote{We are thankful to Dario Bettoni for kindly providing us with his Mathematica notebook performing disformal transformations}. In doing so, we convinced ourselves that, while the disformal transformation does result in extra terms in addition to the standard quadratic kinetic term, some of them are negligible for small $h_{\mu \nu}$, and the rest sum up to a total derivative.

\section{Discussion and conclusions}
\label{sec:Concl}

We have identified the MOND-producing formulation of Superfluid Dark Matter as the Newtonian, low-acceleration limit of a RAQUAL theory written in the Einstein frame. Upon transformation to the Jordan frame, the RAQUAL interaction sector generates two non-minimal curvature couplings: a derivative coupling to the Einstein tensor with the same structure as the BECDM coupling, and a direct coupling of the scalar field to the Ricci scalar. For a constant choice of \(B(\phi,X)\), the former coincides with the BECDM term, while the latter represents a small, linear perturbation of the Einstein--Hilbert sector when \(\phi\ll1\). The interaction sectors of the two theories can therefore be mapped into one another up to this perturbation. Their kinetic sectors, however, remain inequivalent: the non-analytic RAQUAL/SFDM kinetic term cannot be obtained from the canonical quadratic kinetic term of BECDM through a perturbative disformal transformation. SFDM and BECDM are thus closely related at the level of their interactions but are not dynamically equivalent.

This partial mapping nevertheless provides a useful theoretical interpretation of the SFDM phonon--baryon interaction $\mathcal{L}_{\rm int}\propto\pi\rho_b$, see Eq.~\eqref{SFDM-int}, which is introduced phenomenologically to reproduce the baryonic source term in the MOND Poisson equation. The mapping shows that this interaction may be understood as the Einstein-frame representation of a non-minimal coupling between the dark-matter scalar and curvature in the Jordan frame. Such couplings arise naturally in effective descriptions of extended or condensed systems and could, in principle, be generated dynamically on galactic scales. Determining whether an appropriate coupling can be activated only in the dark sector and only in the relevant regime remains an important direction for future work.

The identification with RAQUAL also places the MOND-producing version of SFDM within a broader theoretical context. RAQUAL models that recover MOND face several well-known difficulties: they do not reproduce the required gravitational lensing by galactic halos, their scalar equation of motion can lose hyperbolicity on hypersurfaces where $X=0$, and avoiding MONDian effects in the Solar System generally requires additional fine-tuning \cite{Bruneton_2007}. Because SFDM is intended only as a low-energy description inside superfluid halo cores, it need not automatically inherit every pathology of its relativistic completion. Nevertheless, the loss of hyperbolicity at $X=0$ is potentially relevant even within this restricted domain. Establishing whether $X$ remains nonzero throughout the superfluid core, and whether the resulting initial-value problem is well posed, requires a dedicated analysis. More elaborate RAQUAL constructions can mitigate some of these difficulties by modifying $f(X)$ and introducing genuinely disformal relations with nontrivial functions $A$ and $B$, but existing realizations tend to trade the original problems for ghost instabilities, loss of hyperbolicity within matter, or additional fine-tuning \cite{Bruneton_2007, Bruneton_2009}.

BECDM avoids the pathologies directly associated with the aquadratic RAQUAL kinetic term because its scalar field has a standard quadratic kinetic term. The present mapping nevertheless indicates that BECDM is not expected to ever exactly reproduce MOND in its current form. Matching the derivative non-minimal coupling is only one ingredient: recovering the MOND limit would additionally require either the non-analytic RAQUAL kinetic term or a more elaborate choice of the disformal functions $A(\phi,X)$ and $B(\phi,X)$ \cite{Bruneton_2009}. This does not preclude BECDM from accounting for galactic or cluster phenomenology, since MOND is not the only viable description of galactic dynamics neither it appears to be as successful at cluster scales. Alternative non-minimal couplings, including those formulated consistently in the hydrodynamic framework \cite{Bettoni_2015,Silveravalle_2025,Benetti_2025}, therefore remain worthy of theoretical and observational investigation. Existing results already suggest that such couplings may alleviate the cusp--core problem and provide successful fits to galactic rotation curves \cite{Gandolfi:2021jai,Gandolfi_2022} and consistently with the former, fit observations at the cluster scales~\cite{Gandolfi:2023hwx}.

Finally, empirical studies indicate that the MOND-producing version of SFDM performs somewhat less successfully than MOND itself, largely because the phonon-mediated force operates only within the superfluid core \cite{Hossenfelder_2019, Hossenfelder:2020yko, Mistele:2020qha, Mistele:2022vhh, Mistele:2023wao}. Recent developments have consequently shifted toward simpler scalar-field models with standard kinetic terms and quartic or sextic self-interactions \cite{Berezhiani:2025maf}. The present analysis lends theoretical support to this shift by showing that the earlier MOND-producing formulation is tied to the non-analytic kinetic structure, and hence potentially to the unresolved consistency problems, of RAQUAL. More broadly, the comparison illustrates the value of mapping modified-gravity theories across different frames and formulations: apparent similarities between their interaction sectors need not imply full dynamical equivalence, but they can reveal shared physical mechanisms, transfer known consistency conditions, and clarify which phenomenological results genuinely carry over from one model to another.

\begin{acknowledgments}
We wish to acknowledge Valerio De Luca for an insightful seminar he delivered about Superfluid Dark Matter at SISSA, which served as impetus for this work, and for enlightening discussions afterward. We are furthermore grateful to Farid Thaalba, Emma Albertini and David Maibach for insightful discussions. We are particularly grateful to Dario Bettoni for an enlightening email correspondence and for kindly providing us with his Mathematica notebook which performs disformal transformations.
\end{acknowledgments}

\appendix

\nocite{*}

\bibliography{apssamp}% Produces the bibliography via BibTeX.

@article{Bettoni_2013,
   title={Disformal invariance of second order scalar-tensor theories: Framing the Horndeski action},
   volume={88},
   ISSN={1550-2368},
   url={http://dx.doi.org/10.1103/PhysRevD.88.084020},
   DOI={10.1103/physrevd.88.084020},
   number={8},
   journal={Physical Review D},
   publisher={American Physical Society (APS)},
   author={Bettoni, Dario and Liberati, Stefano},
   year={2013},
   month=oct }

@article{Khoury_2015,
   title={Alternative to particle dark matter},
   volume={91},
   ISSN={1550-2368},
   url={http://dx.doi.org/10.1103/PhysRevD.91.024022},
   DOI={10.1103/physrevd.91.024022},
   number={2},
   journal={Physical Review D},
   publisher={American Physical Society (APS)},
   author={Khoury, Justin},
   year={2015},
   month=jan }

@article{De_Luca_2023,
   title={Superfluid dark matter around black holes},
   volume={2023},
   ISSN={1475-7516},
   url={http://dx.doi.org/10.1088/1475-7516/2023/04/048},
   DOI={10.1088/1475-7516/2023/04/048},
   number={04},
   journal={Journal of Cosmology and Astroparticle Physics},
   publisher={IOP Publishing},
   author={De Luca, Valerio and Khoury, Justin},
   year={2023},
   month=apr, pages={048} }

@article{Berezhiani:2025maf,
    author = "Berezhiani, Lasha and Cintia, Giordano and De Luca, Valerio and Khoury, Justin",
    title = "{Superfluid dark matter}",
    eprint = "2505.23900",
    archivePrefix = "arXiv",
    primaryClass = "astro-ph.CO",
    doi = "10.1016/j.physrep.2026.02.001",
    journal = "Phys. Rept.",
    volume = "1172",
    pages = "1--72",
    year = "2026"
}

@article{Khoury_2022,
   title={Dark Matter Superfluidity},
   ISSN={2590-1990},
   url={http://dx.doi.org/10.21468/SciPostPhysLectNotes.42},
   DOI={10.21468/scipostphyslectnotes.42},
   journal={SciPost Physics Lecture Notes},
   publisher={Stichting SciPost},
   author={Khoury, Justin},
   year={2022},
   month=mar }

@article{Bettoni_2011,
   title={Extended $\Lambda$CDM: generalized non-minimal coupling for dark matter fluids},
   volume={2011},
   ISSN={1475-7516},
   url={http://dx.doi.org/10.1088/1475-7516/2011/11/007},
   DOI={10.1088/1475-7516/2011/11/007},
   number={11},
   journal={Journal of Cosmology and Astroparticle Physics},
   publisher={IOP Publishing},
   author={Bettoni, Dario and Liberati, Stefano and Sindoni, Lorenzo},
   year={2011},
   month=nov, pages={007–007} }

@article{Bettoni_2015,
   title={Dynamics of non-minimally coupled perfect fluids},
   volume={2015},
   ISSN={1475-7516},
   url={http://dx.doi.org/10.1088/1475-7516/2015/08/023},
   DOI={10.1088/1475-7516/2015/08/023},
   number={08},
   journal={Journal of Cosmology and Astroparticle Physics},
   publisher={IOP Publishing},
   author={Bettoni, Dario and Liberati, Stefano},
   year={2015},
   month=aug, pages={023–023} }

@article{Bettoni_2012,
doi = {10.1088/1475-7516/2012/07/027},
url = {https://doi.org/10.1088/1475-7516/2012/07/027},
year = {2012},
month = {jul},
publisher = {},
volume = {2012},
number = {07},
pages = {027},
author = {Dario Bettoni and Valeria Pettorino and Stefano Liberati and Carlo Baccigalupi},
title = {Non-minimally coupled dark matter: effective pressure and structure formation},
journal = {Journal of Cosmology and Astroparticle Physics}
}

@article{Bettoni_2014,
   title={Dark matter as a Bose-Einstein Condensate: the relativistic non-minimally coupled case},
   volume={2014},
   ISSN={1475-7516},
   url={http://dx.doi.org/10.1088/1475-7516/2014/02/004},
   DOI={10.1088/1475-7516/2014/02/004},
   number={02},
   journal={Journal of Cosmology and Astroparticle Physics},
   publisher={IOP Publishing},
   author={Bettoni, Dario and Colombo, Mattia and Liberati, Stefano},
   year={2014},
   month=feb, pages={004–004} }

@article{Bruneton_2009,
   title={Reconciling MOND and dark matter?},
   volume={2009},
   ISSN={1475-7516},
   url={http://dx.doi.org/10.1088/1475-7516/2009/03/021},
   DOI={10.1088/1475-7516/2009/03/021},
   number={03},
   journal={Journal of Cosmology and Astroparticle Physics},
   publisher={IOP Publishing},
   author={Bruneton, Jean-Philippe and Liberati, Stefano and Sindoni, Lorenzo and Famaey, Benoit},
   year={2009},
   month=mar, pages={021–021} }

@article{Sotiriou:2007zu,
    author = "Sotiriou, Thomas P and Faraoni, Valerio and Liberati, Stefano",
    title = "{Theory of gravitation theories: A No-progress report}",
    eprint = "0707.2748",
    archivePrefix = "arXiv",
    primaryClass = "gr-qc",
    doi = "10.1142/S0218271808012097",
    journal = "Int. J. Mod. Phys. D",
    volume = "17",
    pages = "399--423",
    year = "2008"
}

@article{Kuntz:2026vhs,
    author = "Kuntz, Iber{\^e} and Liberati, Stefano",
    title = "{Off-shell equivalence in quantum field theory and gravity}",
    eprint = "2607.12644",
    archivePrefix = "arXiv",
    primaryClass = "hep-th",
    month = "7",
    year = "2026",
    journal=""
}

@article{Bruneton_2007,
  title = {Field-theoretical formulations of MOND-like gravity},
  author = {Bruneton, Jean-Philippe and Esposito-Far\`ese, Gilles},
  journal = {Phys. Rev. D},
  volume = {76},
  issue = {12},
  pages = {124012},
  numpages = {41},
  year = {2007},
  month = {Dec},
  publisher = {American Physical Society},
  doi = {10.1103/PhysRevD.76.124012},
  url = {https://link.aps.org/doi/10.1103/PhysRevD.76.124012}
}

@article{Bekenstein_1993,
   title={Relation between physical and gravitational geometry},
   volume={48},
   ISSN={0556-2821},
   url={http://dx.doi.org/10.1103/PhysRevD.48.3641},
   DOI={10.1103/physrevd.48.3641},
   number={8},
   journal={Physical Review D},
   publisher={American Physical Society (APS)},
   author={Bekenstein, Jacob D.},
   year={1993},
   month=oct, pages={3641–3647} }

@article{Bekenstein:2010pt,
    author = "Bekenstein, Jacob D.",
    title = "{Alternatives to Dark Matter: Modified Gravity as an Alternative to dark Matter}",
    eprint = "1001.3876",
    archivePrefix = "arXiv",
    primaryClass = "astro-ph.CO",
    pages = "99--117",
    month = "1",
    year = "2010",
    journal=""
    
}

@article{Benetti_2025,
   title={Ultra-compact objects of non-minimally coupled dark matter},
   volume={2025},
   ISSN={1475-7516},
   url={http://dx.doi.org/10.1088/1475-7516/2025/03/029},
   DOI={10.1088/1475-7516/2025/03/029},
   number={03},
   journal={Journal of Cosmology and Astroparticle Physics},
   publisher={IOP Publishing},
   author={Benetti, Francesco and Lapi, Andrea and Silveravalle, Samuele and Liberati, Stefano},
   year={2025},
   month=mar, pages={029} }

@article{Silveravalle_2025,
doi = {10.1088/1475-7516/2025/11/067},
url = {https://doi.org/10.1088/1475-7516/2025/11/067},
year = {2025},
month = {nov},
publisher = {IOP Publishing},
volume = {2025},
number = {11},
pages = {067},
author = {Silveravalle, Samuele and Lapi, Andrea and Benetti, Francesco and Liberati, Stefano},
title = {Cosmology with a non-minimally coupled dark matter fluid. Part I. Background evolution},
journal = {Journal of Cosmology and Astroparticle Physics}
}

@article{Gandolfi_2022,
   title={Empirical Evidence of Nonminimally Coupled Dark Matter in the Dynamics of Local Spiral Galaxies?},
   volume={929},
   ISSN={1538-4357},
   url={http://dx.doi.org/10.3847/1538-4357/ac5970},
   DOI={10.3847/1538-4357/ac5970},
   number={1},
   journal={The Astrophysical Journal},
   publisher={American Astronomical Society},
   author={Gandolfi, Giovanni and Lapi, Andrea and Liberati, Stefano},
   year={2022},
   month=apr, pages={48} }

@article{Gandolfi:2023hwx,
    author = "Gandolfi, Giovanni and Haridasu, Balakrishna Sandeep and Liberati, Stefano and Lapi, Andrea",
    title = "{Looking for Traces of Nonminimally Coupled Dark Matter in the X-COP Galaxy Clusters Sample}",
    eprint = "2305.13974",
    archivePrefix = "arXiv",
    primaryClass = "astro-ph.CO",
    doi = "10.3847/1538-4357/acd755",
    journal = "Astrophys. J.",
    volume = "952",
    number = "2",
    pages = "105",
    year = "2023"
}

@article{Flanagan_2004,
    author = "Flanagan, Eanna E.",
    title = "{The Conformal frame freedom in theories of gravitation}",
    eprint = "gr-qc/0403063",
    archivePrefix = "arXiv",
    doi = "10.1088/0264-9381/21/15/N02",
    journal = "Class. Quant. Grav.",
    volume = "21",
    pages = "3817",
    year = "2004"
}

@ARTICLE{Milgrom_1983,
       author = {{Milgrom}, M.},
        title = "{A modification of the Newtonian dynamics as a possible alternative to the hidden mass hypothesis.}",
      journal = {\apj},
         year = 1983,
        month = jul,
       volume = {270},
        pages = {365-370},
          doi = {10.1086/161130},
       adsurl = {https://ui.adsabs.harvard.edu/abs/1983ApJ...270..365M}
}

@article{Famaey_2012,
   title={Modified Newtonian Dynamics (MOND): Observational Phenomenology and Relativistic Extensions},
   volume={15},
   ISSN={1433-8351},
   url={http://dx.doi.org/10.12942/lrr-2012-10},
   DOI={10.12942/lrr-2012-10},
   number={1},
   journal={Living Reviews in Relativity},
   publisher={Springer Science and Business Media LLC},
   author={Famaey, Benoît and McGaugh, Stacy S.},
   year={2012},
   month=Sept }

@article{Milgrom:2014usa,
    author = "Milgrom, Mordehai",
    title = "{MOND theory}",
    eprint = "1404.7661",
    archivePrefix = "arXiv",
    primaryClass = "astro-ph.CO",
    doi = "10.1139/cjp-2014-0211",
    journal = "Can. J. Phys.",
    volume = "93",
    number = "2",
    pages = "107--118",
    year = "2015"
}

@ARTICLE{1984ApJ...286....7B,
       author = {{Bekenstein}, J. and {Milgrom}, M.},
        title = "{Does the missing mass problem signal the breakdown of Newtonian gravity?}",
      journal = {\apj},
         year = 1984,
        month = nov,
       volume = {286},
        pages = {7-14},
          doi = {10.1086/162570},
       adsurl = {https://ui.adsabs.harvard.edu/abs/1984ApJ...286....7B}
}

@article{Barcel__2005,
   title={Analogue Gravity},
   volume={8},
   ISSN={1433-8351},
   url={http://dx.doi.org/10.12942/lrr-2005-12},
   DOI={10.12942/lrr-2005-12},
   number={1},
   journal={Living Reviews in Relativity},
   publisher={Springer Science and Business Media LLC},
   author={Barceló, Carlos and Liberati, Stefano and Visser, Matt},
   year={2005},
   month=Dec }

@article{Zumalac_rregui_2013,
   title={DBI Galileons in the Einstein frame: Local gravity and cosmology},
   volume={87},
   ISSN={1550-2368},
   url={http://dx.doi.org/10.1103/PhysRevD.87.083010},
   DOI={10.1103/physrevd.87.083010},
   number={8},
   journal={Physical Review D},
   publisher={American Physical Society (APS)},
   author={Zumalacárregui, Miguel and Koivisto, Tomi S. and Mota, David F.},
   year={2013},
   month=Apr }

@article{PhysRevD.111.024053,
  title = {Hyperbolicity in scalar-Gauss-Bonnet gravity: A gauge invariant study for spherical evolution},
  author = {Thaalba, Farid and Franchini, Nicola and Bezares, Miguel and Sotiriou, Thomas P.},
  journal = {Phys. Rev. D},
  volume = {111},
  issue = {2},
  pages = {024053},
  numpages = {16},
  year = {2025},
  month = {Jan},
  publisher = {American Physical Society},
  doi = {10.1103/PhysRevD.111.024053},
  url = {https://link.aps.org/doi/10.1103/PhysRevD.111.024053}
}

@article{Mistele:2022vhh,
    author = "Mistele, Tobias and McGaugh, Stacy and Hossenfelder, Sabine",
    title = "{Galactic mass-to-light ratios with superfluid dark matter}",
    eprint = "2201.07282",
    archivePrefix = "arXiv",
    primaryClass = "astro-ph.GA",
    doi = "10.1051/0004-6361/202243216",
    journal = "Astron. Astrophys.",
    volume = "664",
    pages = "A40",
    year = "2022"
}

@article{Hossenfelder_2019,
   title={Strong lensing with superfluid dark matter},
   volume={2019},
   ISSN={1475-7516},
   url={http://dx.doi.org/10.1088/1475-7516/2019/02/001},
   DOI={10.1088/1475-7516/2019/02/001},
   number={02},
   journal={Journal of Cosmology and Astroparticle Physics},
   publisher={IOP Publishing},
   author={Hossenfelder, Sabine and Mistele, Tobias},
   year={2019},
   month=Feb, pages={001–001} }

@article{Mistele:2020qha,
    author = "Mistele, Tobias",
    title = "{Three problems of superfluid dark matter and their solution}",
    eprint = "2009.03003",
    archivePrefix = "arXiv",
    primaryClass = "gr-qc",
    doi = "10.1088/1475-7516/2021/01/025",
    journal = "JCAP",
    volume = "01",
    pages = "025",
    year = "2021"
}

@article{Hossenfelder:2020yko,
    author = "Hossenfelder, Sabine and Mistele, Tobias",
    title = "{The Milky Way{\textquoteright}s rotation curve with superfluid dark matter}",
    eprint = "2003.07324",
    archivePrefix = "arXiv",
    primaryClass = "astro-ph.GA",
    doi = "10.1093/mnras/staa2594",
    journal = "Mon. Not. Roy. Astron. Soc.",
    volume = "498",
    number = "3",
    pages = "3484--3491",
    year = "2020"
}

@article{Berezhiani:2023vlo,
    author = "Berezhiani, Lasha and Cintia, Giordano and De Luca, Valerio and Khoury, Justin",
    title = "{Dynamical friction in dark matter superfluids: The evolution of black hole binaries}",
    eprint = "2311.07672",
    archivePrefix = "arXiv",
    primaryClass = "astro-ph.CO",
    doi = "10.1088/1475-7516/2024/06/024",
    journal = "JCAP",
    volume = "06",
    pages = "024",
    year = "2024"
}

@article{Berezhiani:2022buv,
    author = "Berezhiani, Lasha and Cintia, Giordano and Khoury, Justin",
    title = "{Thermalization, fragmentation, and tidal disruption: The complex galactic dynamics of dark matter superfluidity}",
    eprint = "2212.10577",
    archivePrefix = "arXiv",
    primaryClass = "astro-ph.CO",
    doi = "10.1103/PhysRevD.107.123010",
    journal = "Phys. Rev. D",
    volume = "107",
    number = "12",
    pages = "123010",
    year = "2023"
}

@article{Berezhiani:2021rjs,
    author = "Berezhiani, Lasha and Cintia, Giordano and Warkentin, Max",
    title = "{Core fragmentation in simplest superfluid dark matter scenario}",
    eprint = "2101.08117",
    archivePrefix = "arXiv",
    primaryClass = "astro-ph.CO",
    doi = "10.1016/j.physletb.2021.136422",
    journal = "Phys. Lett. B",
    volume = "819",
    pages = "136422",
    year = "2021"
}

@article{Berezhiani:2020umi,
    author = "Berezhiani, Lasha",
    title = "{On effective theory of superfluid phonons}",
    eprint = "2001.08696",
    archivePrefix = "arXiv",
    primaryClass = "hep-th",
    doi = "10.1016/j.physletb.2020.135451",
    journal = "Phys. Lett. B",
    volume = "805",
    pages = "135451",
    year = "2020"
}

@article{Berezhiani:2019pzd,
    author = "Berezhiani, Lasha and Elder, Benjamin and Khoury, Justin",
    title = "{Dynamical Friction in Superfluids}",
    eprint = "1905.09297",
    archivePrefix = "arXiv",
    primaryClass = "hep-ph",
    doi = "10.1088/1475-7516/2019/10/074",
    journal = "JCAP",
    volume = "10",
    pages = "074",
    year = "2019"
}

@article{Berezhiani:2018oxf,
    author = "Berezhiani, Lasha and Khoury, Justin",
    title = "{Emergent long-range interactions in Bose-Einstein Condensates}",
    eprint = "1812.09332",
    archivePrefix = "arXiv",
    primaryClass = "hep-th",
    doi = "10.1103/PhysRevD.99.076003",
    journal = "Phys. Rev. D",
    volume = "99",
    number = "7",
    pages = "076003",
    year = "2019"
}

@article{Berezhiani:2017tth,
    author = "Berezhiani, Lasha and Famaey, Benoit and Khoury, Justin",
    title = "{Phenomenological consequences of superfluid dark matter with baryon-phonon coupling}",
    eprint = "1711.05748",
    archivePrefix = "arXiv",
    primaryClass = "astro-ph.CO",
    doi = "10.1088/1475-7516/2018/09/021",
    journal = "JCAP",
    volume = "09",
    pages = "021",
    year = "2018"
}

@article{Gandolfi:2021jai,
    author = "Gandolfi, Giovanni and Lapi, Andrea and Liberati, Stefano",
    title = "{Self-gravitating Equilibria of Non-minimally Coupled Dark Matter Halos}",
    eprint = "2102.03873",
    archivePrefix = "arXiv",
    primaryClass = "astro-ph.CO",
    doi = "10.3847/1538-4357/abe460",
    journal = "Astrophys. J.",
    volume = "910",
    number = "1",
    pages = "76",
    year = "2021"
}

@article{Berezhiani:2015bqa,
    author = "Berezhiani, Lasha and Khoury, Justin",
    title = "{Theory of dark matter superfluidity}",
    eprint = "1507.01019",
    archivePrefix = "arXiv",
    primaryClass = "astro-ph.CO",
    doi = "10.1103/PhysRevD.92.103510",
    journal = "Phys. Rev. D",
    volume = "92",
    pages = "103510",
    year = "2015"
}

@article{Berezhiani:2015pia,
    author = "Berezhiani, Lasha and Khoury, Justin",
    title = "{Dark Matter Superfluidity and Galactic Dynamics}",
    eprint = "1506.07877",
    archivePrefix = "arXiv",
    primaryClass = "astro-ph.CO",
    doi = "10.1016/j.physletb.2015.12.054",
    journal = "Phys. Lett. B",
    volume = "753",
    pages = "639--643",
    year = "2016"
}

@article{Mistele:2023wao,
    author = "Mistele, Tobias and McGaugh, Stacy and Hossenfelder, Sabine",
    title = "{Superfluid dark matter in tension with weak gravitational lensing data}",
    eprint = "2303.08560",
    archivePrefix = "arXiv",
    primaryClass = "astro-ph.GA",
    doi = "10.1088/1475-7516/2023/09/004",
    journal = "JCAP",
    volume = "09",
    pages = "004",
    year = "2023"
}

\end{document}